\documentclass[aps,prb,twocolumn,10pt,floatfix,eqsecnum,longbibliography,nofootinbib,showkeys,reprint,nobibnotes]{revtex4-2}

\usepackage{amsmath}
\usepackage{amssymb}
\usepackage{amsfonts}
\usepackage{tablefootnote}
\usepackage{graphicx}
\usepackage{color}
\usepackage[dvipsnames]{xcolor}
\usepackage{bbm}
\usepackage[breaklinks=true,colorlinks=true,linkcolor=blue,urlcolor=blue,citecolor=blue]{hyperref}
\usepackage{array}
\usepackage{booktabs}
\usepackage{mathptmx}
\usepackage{placeins}

\DeclareMathAlphabet{\mathcal}{OMS}{cmsy}{m}{n}
\newcolumntype{P}[1]{>{\centering\arraybackslash}p{#1}}
\newcolumntype{M}[1]{>{\centering\arraybackslash}m{#1}}

\begin{document}

\title{Vortex frequency locking and Shapiro steps in superconductor open nanotubes}

\author{Igor Bogush\textsuperscript{1,2}}
\email{igori.bogus@tu-braunschweig.de}
\author{Vladimir M. Fomin\textsuperscript{3,4}}
\author{Oleksandr V. Dobrovolskiy\textsuperscript{1,2}}

\affiliation{
\textsuperscript{1}Cryogenic Quantum Electronics, Institute for Electrical Measurement Science and Fundamental Electrical Engineering, Technische Universit\"at Braunschweig, Hans-Sommer-Str. 66, D-38106 Braunschweig, Germany\\
\textsuperscript{2}Laboratory for Emerging Nanometrology (LENA), Technische Universit\"at Braunschweig, Langer Kamp 6A-B, D-38106 Braunschweig,  Germany\\
\textsuperscript{3}Institute for Emerging Electronic Technologies, Leibniz IFW Dresden, Helmholtzstra\ss e 20, D-01069 Dresden, Germany\\
\textsuperscript{4}Moldova State University, str. Alexei Mateevici 60, MD-2009 Chişinău, Republic of Moldova
}

\begin{abstract}
The movement of magnetic flux quanta (Abrikosov vortices) in superconductors leads to dissipation and is influenced by various ordering effects arising from vortex-vortex, vortex-defect, and vortex-edge interactions. Under combined dc and ac stimuli, when the distance traveled by fluxons during an ac cycle corresponds to an integer multiple of the vortex lattice period, the superconductor's current-voltage ($I$-$V$) curve displays synchronization (Shapiro) steps. However, in planar constrictions, frequency-locking effects rely on a perfectly ordered vortex lattice and are typically observed when periodic vortex pinning arrays dominate over intrinsic uncorrelated disorder. Here, we propose 3D superconducting open nanotubes as systems free of periodic disorder, where the $I$-$V$ curves are expected to display pronounced Shapiro steps. Using the time-dependent Ginzburg-Landau equation, we attribute the predicted effect to a reduction in the dimensionality of vortex motion. Namely, rolling a planar film into a tube causes the 2D vortex array, which initially moves throughout the film, to evolve into quasi-1D vortex chains that are restricted to areas where the normal component of the magnetic field is near its maximum. The discussed effects are relevant for superconducting devices, where vortex nucleation frequency and voltage stabilization by an external ac stimulus can enhance their operation.
\end{abstract}

\maketitle
\counterwithout{equation}{section}

\section{Introduction}

Moving magnetic flux quanta (Abrikosov vortices or fluxons) in superconductors give rise to dissipation and finite resistance. When the driving Lorentz-like force exerted on vortices by a transport current exceeds the pinning force, the vortices nucleate at one edge of the superconducting constriction, move across it, and denucleate at the opposite edge\,\cite{Bra95rpp}. In addition, the vortices experience vortex-vortex, vortex-defect and vortex-edge interactions\,\cite{Zel94prl,Emb17nac,Mik21prb}, which can lead to vortex arrangements that exhibit or lack a long-range order. For instance, the motion of a perfectly ordered vortex lattice leads to electromagnetic emission\,\cite{Kul66spj,Bul06prl,Dob18apl,Dob18nac} and, in return, its exposure to an ac stimulus leads to a \emph{frequency-locking effect} (FLE)\,\cite{Mar75ssc,Fio71prl}. This effect manifests as plateaus (Shapiro steps) in the superconductor's current-voltage ($I$-$V$) curve. However, a long-range order in the vortex lattice is difficult to achieve because of a nonuniformity of the transport current\,\cite{Emb17nac}, variation in the strength of the edge barrier\,\cite{Gri10prb}, and the presence of intrinsic disorder. The latter effect can be partially overcome by using periodic pinning site arrays\,\cite{Loo99prb,Von09prb} or stroboscopic synchronization of the vortex motion\,\cite{Jel16nsr}. Yet, these approaches imply nanostructuring or an optical beam impact over the \emph{entire area} of the superconductor.

Superconductor open nanotubes represent a remarkable class of systems where intriguing phenomena in the vortex dynamics emerge due to the 3D geometry\,\cite{Fom12nal,Fom22apl}. For instance, when an external magnetic field $\mathbf{B}$ is applied in a direction that is not parallel to the nanotube axis, its normal $B_\mathrm{n}$ and tangential $B_\mathrm{t}$ components at the sample surface are strongly inhomogeneous. The vortices are confined to move in the quasi-1D tube areas where $B_\mathrm{n}$ is close to maximum\,\cite{bogush2024microwave}. A reduction in the dimensionality of vortex motion from 2D in planar films to quasi-1D in open superconductor tubes enhances the vortex ordering, lifts the requirement for entire-area external impacts, and should lead to the appearance of pronounced Shapiro steps in their $I$-$V$ curves. 

Originally, Shapiro steps were observed for microwave-irradiated Josephson junctions\,\cite{sha64rmp} in which a phase-locking effect (in addition to the FLE) is responsible for their appearance\,\cite{sul70aip}. Since then, Shapiro steps were observed for various systems mimicking the behavior of Josephson junctions, such as superconducting bridges\,\cite{day67physrev,yuz99pcs,naw13prl,ust22jetp}, vortex lattices interacting with periodic pinning\,\cite{rei00prb,rei99prl,van99prb}, and phase-slip lines \cite{dmi07sst,siv03prl}. Shapiro steps also occur for ac-driven skyrmions\,\cite{Rei15prb}, coupled ultrasound and charge density waves\,\cite{mori23apl}, and spin waves interacting with moving vortices\,\cite{Dob21arx}. At the same time, these systems have been presented by planar structures so far.

Recent advancements in the nanofabrication have enabled the creation of 3D superconductor nanoarchitectures with intricate shapes and curved geometries\,\cite{Fom21boo,Mak22adm}. Direct writing using focused ion and electron beams \cite{Fer20mat,Orus2021,Hof23arx} and strain-relaxation-driven self-rolling\,\cite{Thurmer08,Thurmer10,Loe19acs} allow for the fabrication of cylindrical and helical superconductors, appealing for experimental studies and theoretical investigations of their magneto-transport properties. Although time- and spatially-resolved experimental exploration of the vortex dynamics in 3D nanoarchitectures is challenging\,\cite{Fom22apl,Mak22adm}, the analysis of volume-integrated observables, such as the voltage and its frequency spectrum offers a viable approach for probing the vortex dynamics in these complex systems.
\begin{figure*}[t]
    \centering
    \includegraphics[width=0.95\textwidth]{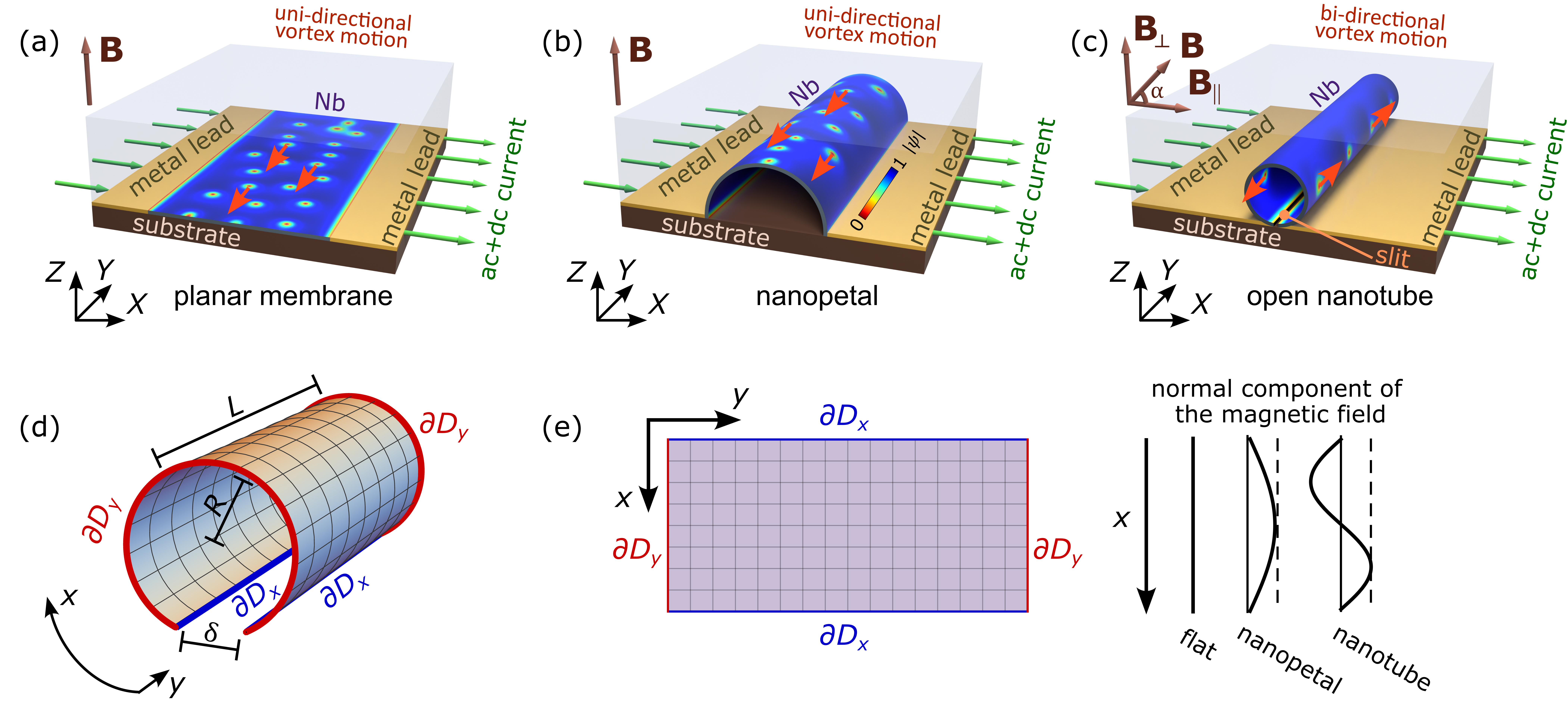}
    \caption{3D representation of the planar membrane (a), nanopetal (b), and open nanotube (c). The distribution of the order parameter magnitude is overlaid on the superconducting structure (with $\alpha = 0$ for the nanotube). (d) The geometrical model of the nanotube and (e) the unwrapped view of the nanotube/nanopetal surface with a schematic representation of the profile of the normal component of the magnetic induction. Metal--superconductor contacts that provide the transport current density $j_\text{tr}$ are located at the slit edges and correspond to the $\partial D_x$ boundaries.}
    \label{fig:1}
\end{figure*}

Here, we predict a pronounced FLE for the vortex motion in (dc+ac)-driven superconductor open nanotubes. The nanotubes do not contain artificial pinning site arrays, are free from edge defects, and are not exposed to any further entire-volume impacts. The synchronization effect originates solely from the enhancement in the ordering of the moving vortices, caused by their confinement to the quasi-1D tube areas where $B_\mathrm{n}$ is close to maximum. The FLE leads to Shapiro steps in the $I$-$V$ curve of the superconductor and is also revealed to stabilize the average voltage against the magnetic-field tilting. Our findings highlight the effect of geometry-induced vortex confinement and the transition from the 2D to a quasi-1D vortex dynamics enabled by 3D nanoarchitectures.

\section{Results}

\subsection{Geometry}

Our study focuses on superconducting nanostructures made of thin films and shaped into three different geometries. These are a planar membrane, a nanopetal, and an open nanotube, Fig.\,\ref{fig:1}(a-c), respectively. All structures are assumed to be made from a film with a thickness of $50$\,nm\,\cite{Loe19acs} and to have superconducting parameters typical of Nb films. The planar membrane has dimensions $L \times W$ while the nanotube and the nanopetal have length $L$, curvature radius $R$, and arc-length $W$. In addition, the nanotube has a small slit of size $\delta$. The geometrical parameters of the considered structures are summarized in Table\,\ref{table:geometry}. 
\begin{table}[b!]
    \centering
    \caption{Geometrical parameters of the nanoarchitectures.}
    \begin{tabular}{P{3.4cm}P{1.5cm}P{1.6cm}P{1.6cm}}
    \hline
    &
    \textbf{Planar}
    &
    \textbf{Nanopetal}
    &
    \textbf{Nanotube}
    \\
    \midrule
    Length $L$, nm
    & $5000$
    & $5000$
    & $5000$
    \\
    Curvature radius $R$, nm
    & $\infty$
    & $780$
    & $390$
    \\
    Arc-len./width $W$, nm
    & $2390$
    & $2390$
    & $2390$
    \\
    Slit $\delta$, nm
    & --
    & --
    & $60$
    \\
     Thickness $d$, nm
    & $50$
    & $50$
    & $50$
    \\
    \midrule
    \bottomrule
    \end{tabular}
    \label{table:geometry}
 \end{table}
 The width of the flat membrane and the arc-length of the nanopetal is equal to the arc-length of the nanotube $W_\mathrm{planar} = W_\mathrm{petal} = 2\pi\cdot R_\mathrm{tube}$. 

The normally conducting metal layer is interrupted underneath the nanostructure, forming two leads connected to the opposite banks of the nanostructures. A transport current of density $j_\text{tr}$ is applied to the metal leads and the experimental observable is the average voltage $U$ between them and the voltage frequency spectrum $f_\mathrm{U}$. The applied current is the sum of a dc component and a high-frequency ac component. The nanostructures are exposed to the magnetic field $\mathbf{B}$. The magnetic field can be tilted in the plane perpendicular to the tube axis, between the parallel-to-substrate orientation $\mathrm{B}_\parallel$ and the perpendicular-to-substrate orientation $\mathrm{B}_\perp$. The mathematical model of the nanotube geometry is depicted in Fig.\,\ref{fig:1}(d). The coordinates $x$ and $y$ parameterizing the surfaces of all three nanostructures are orthonormal, Fig.\,\ref{fig:1}(e). The modeling is performed relying upon a numerical solution of the time-dependent Ginzburg-Landau (TDGL) equation \cite{Fom12nal,Fom22nsr}, as detailed in Appendix\,\ref{app:model}.

\subsection{Vortex dynamics}

\begin{figure}[t]
    \centering    
    \includegraphics[width=6.6cm]{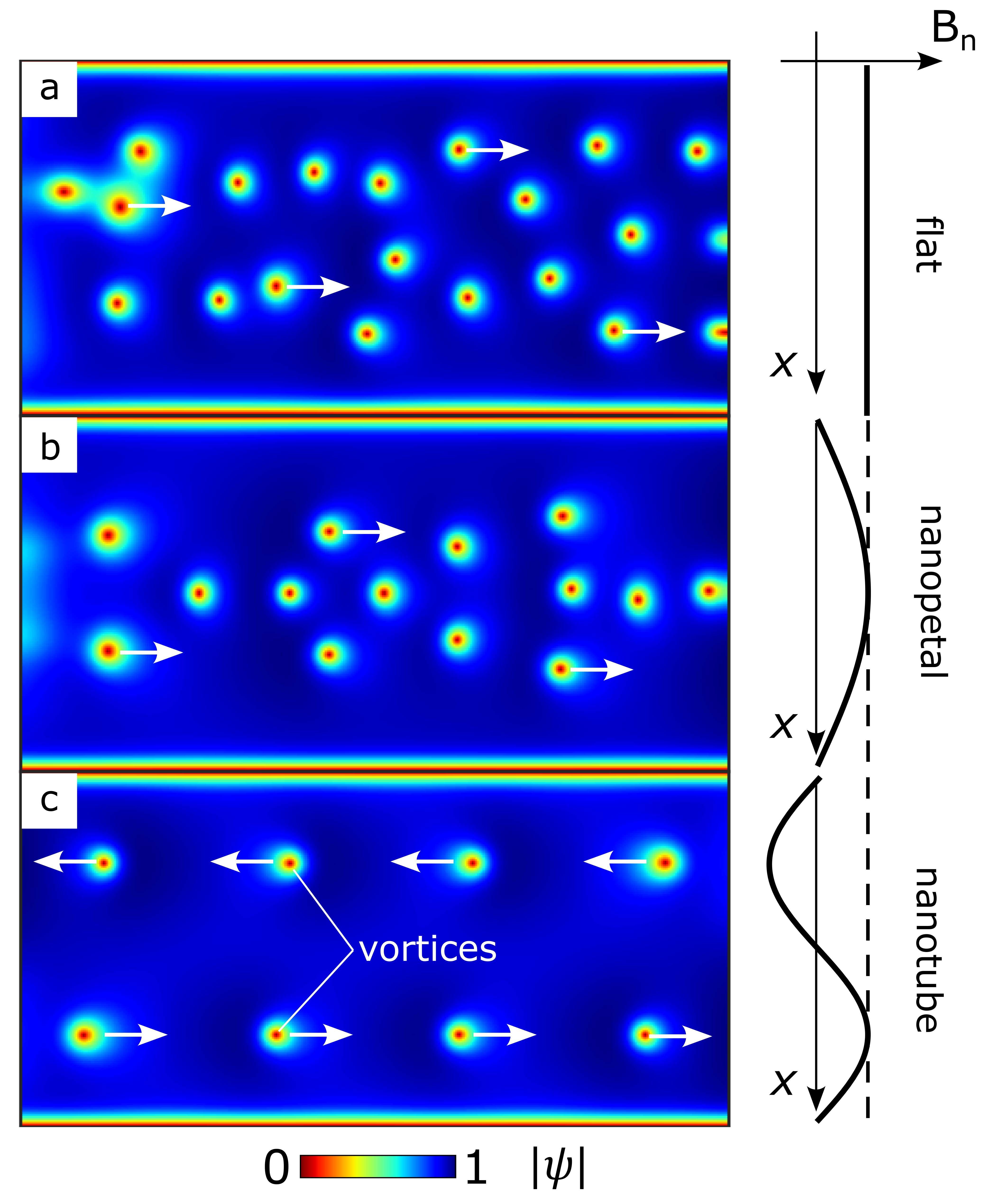}
    \caption{Magnitude of the superconducting order parameter $|\psi|$ in the planar membrane (a), nanopetal (b), and open nanotube (c) at $B=5.2$\,mT and $j_\mathrm{tr}=17$\,GA/m$^2$. The profiles of the normal component of the magnetic field are shown on the right. In the 3D nanostructures, the vortices tend to move in the areas where $B_\mathrm{n}$ is close to maximum.}
    \label{fig:2}
\end{figure}

When the superconductor membrane is exposed to a magnetic field and a transport current exceeding the depinning current, the vortices nucleate at its free edge $\partial D_y$, traverse the sample, and denucleate at the opposite edge as shown in Fig.\,\ref{fig:2}. 
While vortices in the planar membrane move in the same direction, those in the open nanotube and the nanopetal can move either in one direction or in opposite directions, depending on the direction and the magnitude of the magnetic field. Specifically, in low magnetic fields, vortices move in one direction when the magnetic field is perpendicular to the substrate. By contrast, the vortices move in opposite directions in each half of the nanotube and the nanopetal when the magnetic field is parallel to the substrate plane\,\cite{bogush2024steering}. The profiles of the normal component $B_\mathrm{n}(x)$ shown on the right of Fig.\,\ref{fig:2} suggest that in the 3D nanostructures, the vortices tend to move in the areas where $|B_\mathrm{n}|$ is close to maximum\,\cite{Fom12nal}.

While moving vortices can form distinctive patterns, exhibiting periodic or non-periodic behavior\,\cite{bogush2024microwave}, in both cases, one can introduce the {\it vortex nucleation frequency} as a number of vortices nucleated at a given edge per unit time and averaged over many nucleation events. If the vortex motion is regular and periodic, the nucleation frequency appears as a sharp line $f_1$ in the voltage spectrum with the higher harmonics $f_n=nf_1$, where $n\geq2$ is an integer number \cite{bogush2024microwave}. The nucleation frequency $f_1$ depends on the system parameters (strength of the edge barrier, vortex core profile) and increases with the increase of both, the transport current and the magnetic field\,\cite{Fom12nal,bogush2024microwave}. In the nanotube, $f_1$ reflects the number of vortices nucleated in only one half-tube per unit of time.

\subsection{Frequency-locking effect}

An FLE occurs when two oscillating systems synchronize their frequencies, maintaining a constant ratio over time. In the case under consideration, one system is a vortex chain generating dc and ac voltage, while the other one is the combined (dc+ac) transport current of density
\begin{equation}\label{eq:j_modulation}
    j_\mathrm{tr} = j_0 + j_1 \sin (2\pi f_\mathrm{j}  t),
\end{equation}
where $j_0$ is the dc current density, $j_1$ is the ac density amplitude and $f_j$ is the modulation frequency. The effects of moderately and strongly modulated transport currents ($j_1 /j_0 \geq 0.3$) in open superconductor nanotubes were investigated in Ref.\,\cite{Fom22nsr}. Here, we analyze the effects of a weakly modulated transport current ($j_1 \ll j_0$) on the average voltage $U$ and its frequency spectrum $f_\mathrm{U}$ in the regimes of periodic vortex motion without transitional processes between the normal and superconducting states within each ac period. 

In general, the modulation frequency $f_{j}$ can combine with other internal frequencies (e.\,g., $f_1$, $f_2$, etc.) in two ways: perturbatively, creating combinational frequencies (e.\,g., $f_{1} \pm \left|f_{1}-f_\text{j}\right|$) or non-perturbatively, manifesting a resonance. Although the obtained spectra exhibit both perturbative and non-perturbative effects, here we focus on the resonance effects leading to the frequency locking and fractional Shapiro steps. In what follows, we consider the nanotube in a magnetic field parallel to the substrate plane, shown in Fig.\,\ref{fig:1}(c) as $\mathbf{B}_\parallel$, unless stated otherwise. 

\begin{figure}[t]
    \centering
    \includegraphics[scale=0.48]{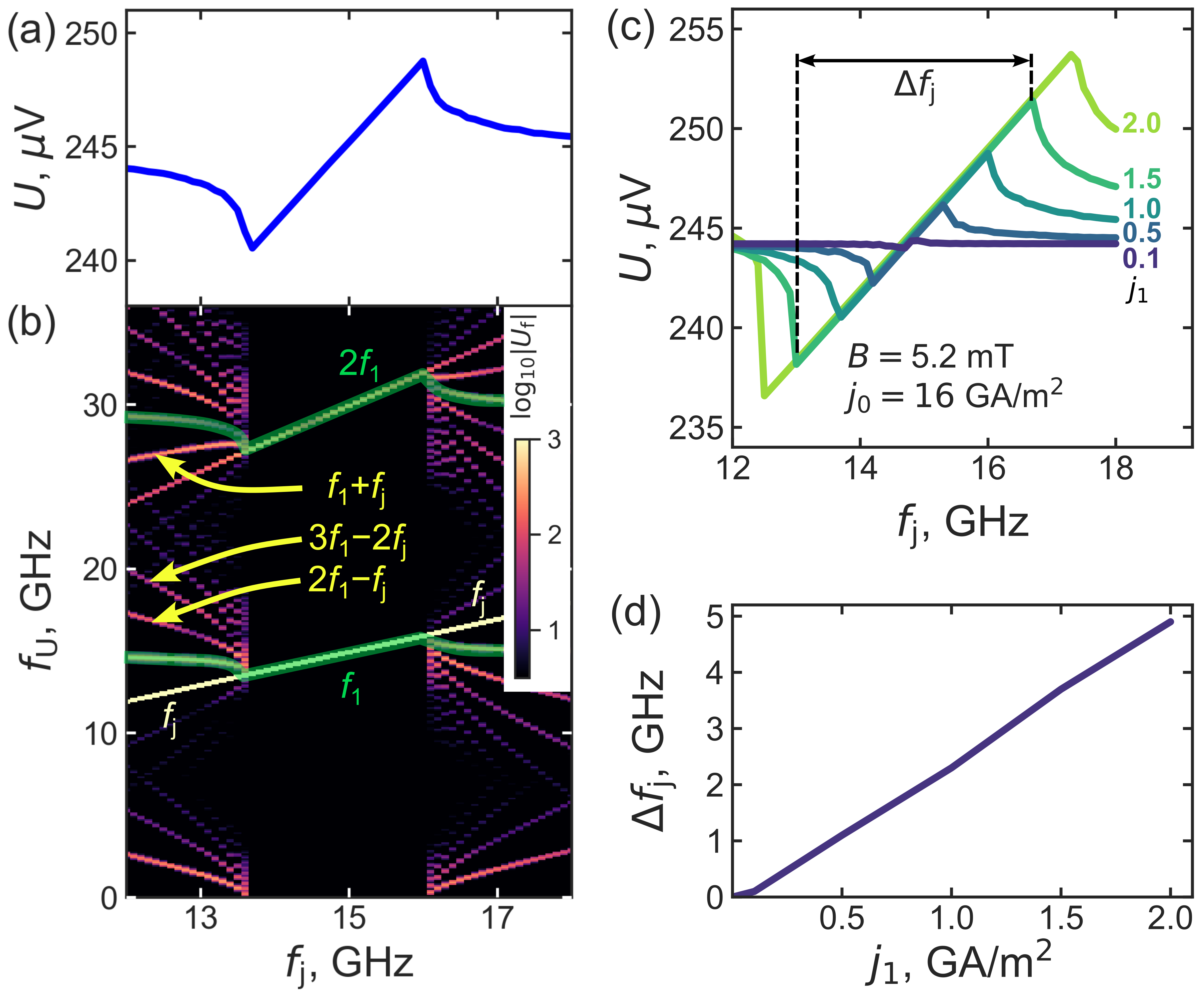}
    \caption{FLEs for the nanotube: Voltage (a) and its spectrum (b) as a function of the modulation frequency $f_\text{j}$ at $B=5.2\text{ mT}$, $j_0=16\text{ GA/m}^2$ and $j_1=1\text{ GA/m}^2$. The green curves highlight the nucleation frequency $f_1$ and its higher harmonic $f_2=2f_1$. The yellow arrows indicate some of the combinational frequencies. (c) Voltage as a function of the modulation frequency $f_\text{j}$ for a series of ac amplitudes $j_1$ and (d) the corresponding FL-range as a function of $j_1$.}
    \label{fig:3}
\end{figure}

Figure \ref{fig:3}(a,b) shows the evolution of $U$ and $f_\mathrm{U}$ as a function of the modulation frequency $f_\text{j}$ of the transport current. The Fourier transformation is used to calculate amplitudes $|U_\text{f}|$ of the wave with frequency $f_\mathrm{U}$ in the voltage spectrum, Fig.\,\ref{fig:3}(b). The vortex nucleation frequency $f_1$ is highlighted by the lower green curve, and its second harmonic $f_2=2f_1$ is highlighted by the upper green curve. The pronounced inclined straight line corresponds to the modulation frequency $f_U=f_\mathrm{j}$ along with its higher harmonics $f_U=2f_\mathrm{j},\,3f_\mathrm{j},$ etc. The voltage spectrum manifests a comb of frequencies below $f_\text{j}\lesssim$ 14 GHz and above $f_\text{j}\gtrsim$ 16 GHz. Each frequency in the comb is a linear combination of the nucleation frequency and the modulation frequency with integer coefficients $n f_1 + m f_\text{j}$, for some integers $n$ and $m$. In the range {$14$ GHz $\lesssim f_\text{j} \lesssim 16$ GHz}, there are only straight lines corresponding to the modulation frequency $f_\text{j}$ and its higher harmonic. In this case, the nucleation frequency is locked by the modulation frequency $f_1=f_\text{j}$ constituting the FLE. We will refer to the set of modulation frequencies where the FLE occurs as the \textit{range of frequency-locking frequencies} or \textit{FL-range}. When the modulation frequency is close to the FL-range, but not in it, the nucleation frequency $f_1$ gravitates to the modulation frequency $f_\text{j}$. The change of the nucleation frequency affects the average voltage $U$ in Fig.\,\ref{fig:3}(a). Outside the FL-range the average voltage decreases, and inside the FL-range it increases, resembling the picture of the anomalous dispersion in optics. A linear growth of the average voltage with the modulation frequency in the frequency-locking regime preserves in a wide range of the modulation amplitudes, as shown in Fig.\, \ref{fig:3}(c). The width of FL-range grows linearly with the amplitude $j_1$, Fig.\,\ref{fig:3}(d). The difference between the highest and lowest average voltage as a function of the modulation frequency $f_\mathrm{j}$ is $17\,\mu$V for the modulation amplitude 2.0 GA/m${}^2$, and it is $8\,\mu$V for 1.0 GA/m${}^2$, what is accessible for experimental verification.

\subsection{Shapiro steps}
\begin{figure}[t]
    \centering
    \includegraphics[width=8.6cm]{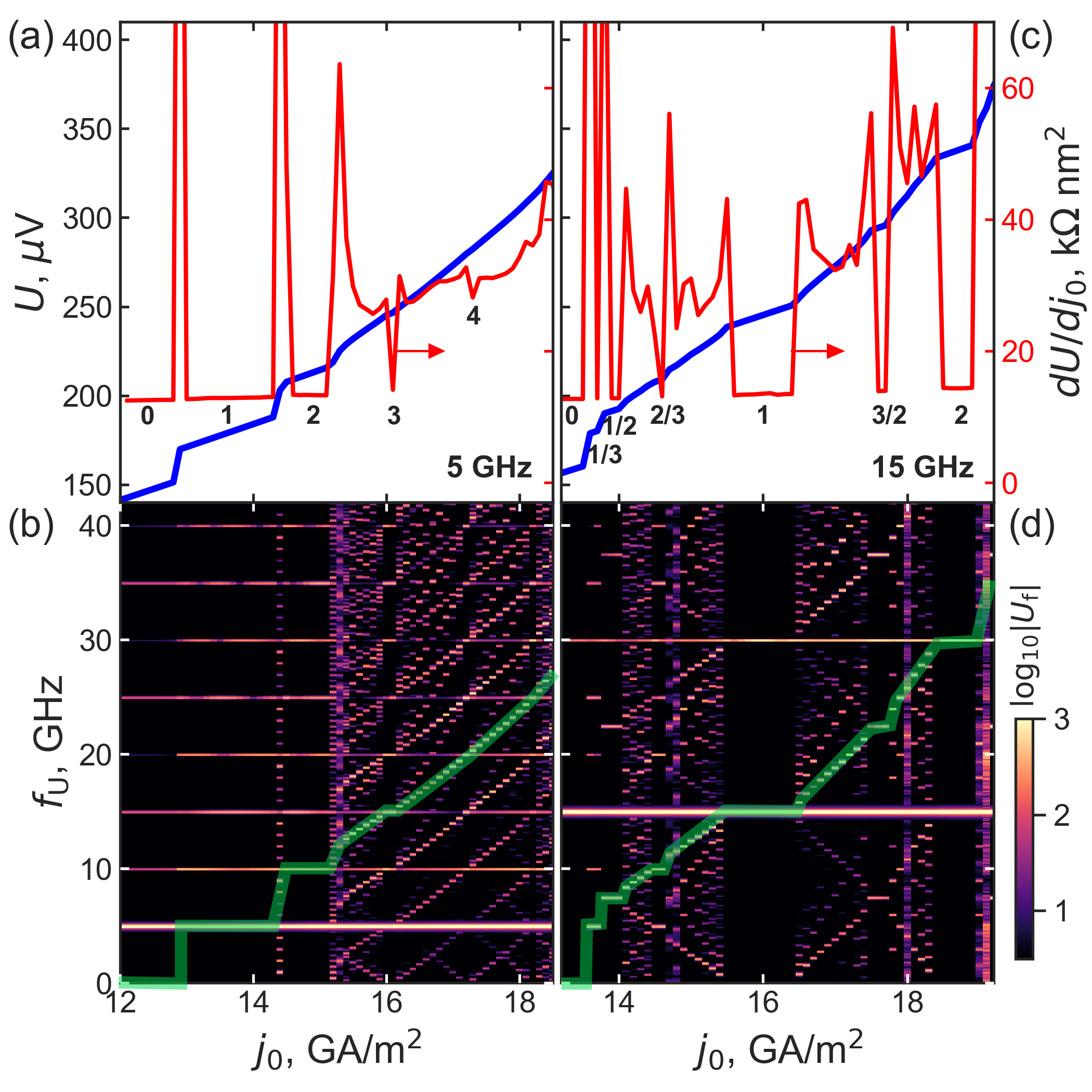}
    \caption{(a,c) Average voltage and (b,d) voltage spectrum as a function of the transport current density in the nanotube with $B=$ 5.2 mT, $j_1=$ 1.5 GA/m${}^2$, for modulation frequency (a,b) $f_\text{j}=$ 5 GHz and (c,d) $f_\text{j}=$ 15 GHz. Integer and fractional order numbers $n$ of Shapiro steps are indicated in the $I$-$V$ curve. The vortex nucleation frequency $f_1$ is highlighted by the green lines.}
    \label{fig:4}
\end{figure}

Figure\,\ref{fig:4} shows the $U(j_0)$ curve for a modulation frequency of $5$ GHz (a,b) and $15$\,GHz (c,d) and an ac density amplitude of $1.5$\,GA/m$^2$. For the modulation frequency of $5$\,GHz, the line $f_\mathrm{U}=5$\,GHz corresponds to the modulation frequency and $f_\mathrm{U}=10,\,15$\,GHz etc. to its higher harmonics. For both frequencies, vortices start nucleating and moving in the nanotube at about $\sim$13 GA/m$^2$, Fig.\,\ref{fig:4}(b,d). Each plateau in the $I$-$V$ curves is characterized by a minimum in the differential resistivity, lying at about $14$\,k$\Omega$nm$^2$. For a lower frequency of $5$\,GHz, the Shapiro steps appear when the nucleation frequency is an integer multiple $n$ of the modulation frequency. In contrast, for the higher frequency of $15$\,GHz, the steps emerge when the nucleation frequencies are rational multiples $n=p/q$ of the modulation frequency, where $p$ and $q$ are some small mutually prime integers. We will refer to the $n$-th Shapiro step with the order number $n$ when the nucleation frequency is locked by the value $f_1 = n f_\mathrm{j}$ in some range of the dc transport current. Thus, both regimes of $5$\,GHz and $15$\,GHz exhibit \emph{integer} Shapiro steps, while the curve for $15$\,GHz additionally displays \emph{fractional} Shapiro steps. As seen in Fig.\,\ref{fig:4}(c), integer steps are wider than the fractional ones.

\begin{figure}
    \centering
    \includegraphics[scale=0.55]{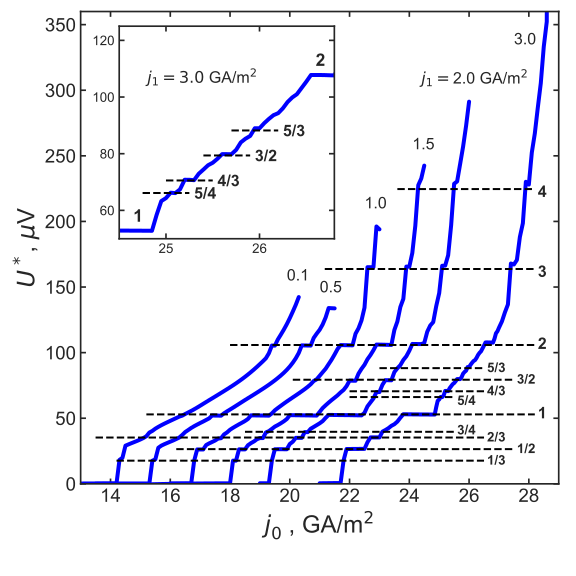}
    \caption{Fractional and integer Shapiro steps in the $I$-$V$ curves of the nanotube at $B=$ 5.2 mT for a series of  modulation amplitudes $j_1$. The order number of a Shapiro step is indicated near the dashed lines. The $I$-$V$ curves are shifted to the right relative to each other for clarity. 
    The inset depicts the enlarged section of the $I$-$V$ curve for $j_1=3$\,GA/m$^2$.}
    \label{fig:5}
\end{figure}
Figure\,\ref{fig:5} shows the $I$-$V$ curves for the nanotube exposed to a magnetic field of $5.2$\,mT at $f_\mathrm{j} = 15$\,GHz. Each curve corresponds to a fixed amplitude of the transport current density modulation. Due to the boundary conditions given by Eqs.\,\eqref{eq:boundaries}, the voltage $U$ includes a contribution of the contact voltage $U_\mathrm{c}$. As a result of the nonlinearity of the TDGL equation, the contact voltage increases nonlinearly with an increase in the transport current and can be approximated with a quadratic function within the considered range of transport currents. Therefore, a quadratic function $U_\mathrm{c} = a_0 + a_1 j_0 + a_2 j_0^2$ has been subtracted from the curves $U^* = U - U_\mathrm{c}$ in order to obtain flat horizontal plateaus at each Shapiro step. The coefficients $a_i$ were fitted to obtain the voltage at $n=2$ twice larger than the voltage at $n=1$, resulting in the following values of $a_0 = 24.88$, $a_1 = 7.22$ and $a_2 = 0.21$, where the voltage and current density are measured in $\mu\text{V}$ and $\text{GA/m}^2$, respectively. 

Note that both integer and fractional Shapiro steps are present in the $I$-$V$ curves in Fig.\,\ref{fig:5}. Fractional Shapiro steps are prevalent in the curves for larger modulation amplitudes. The steps with the same $n$ occur at the same voltages for all curves. The voltages for the Shapiro steps are approximately proportional to the step's number $U^* \approx n U_0$ with $U_0\approx52.9\,\mu$V. An enlarged part of the $I$-$V$ curve for $j_1=$3 GA/m$^2$ is shown in the inset of Fig.\,\ref{fig:5}. Between $n=1$ and $n=2$, it exhibits a large number of fractional Shapiro steps.

\subsection{Tilted magnetic field}

In the absence of an ac modulation, a slight change of the magnetic field direction is known to lead to a change in the vortex nucleation frequency\,\cite{bogush2024steering}. The revealed FLE can therefore be used for the enhancement of the robustness of the vortex motion under a variation of the magnetic field direction. To demonstrate such robustness, we show that a tilt of the magnetic field up to an angle $\alpha$ of a few degrees does not change the nucleation frequency locked to $f_\mathrm{j}$. 

\begin{figure}[t]
    \centering
    \includegraphics[scale=0.55]{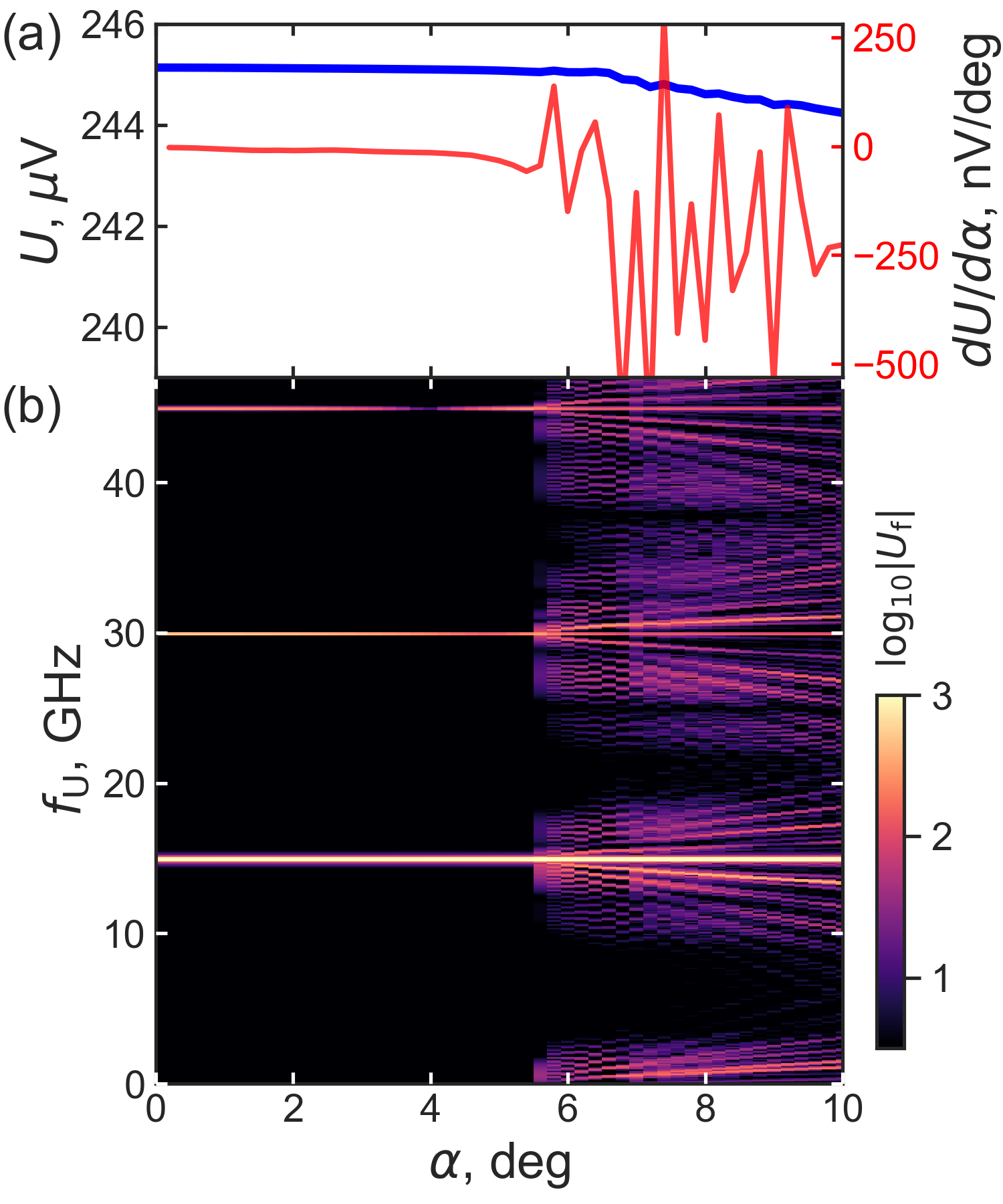}
    \caption{(a) Average voltage, its derivative, and (b) the voltage spectrum as a function of the magnetic field tilt angle $\alpha$. Other parameters are $B=5.2$ mT, $j_0=16.0$ GA/m${}^2$, $j_1=1.0$ GA/m${}^2$, and $f_\text{j}=15$\,GHz.}
    \label{fig:6}
\end{figure}

Figure\,\ref{fig:6} shows the evolution of the average voltage and the voltage frequency spectrum in the open nanotube with an increase in the magnetic field tilt angle $\alpha$ at the modulation frequency of $15$\,GHz, $j_1=1.0$\,GA/m$^2$, and $B=5.2$\,mT. Up to approximately $\alpha\sim5^\circ$, the average voltage and the vortex nucleation frequency remain constant. With an increase in the tilt angle, the nucleation frequencies in the half-tubes split, creating a frequency comb. In the same open nanotube without ac modulation, this splitting occurs at much smaller $\alpha$\,\cite{bogush2024steering}. Accordingly, for the $1$\,GA/m$^2$ modulation of the transport current density, the nucleation frequency $f_1$ can be kept constant up to a field tilt angle of $\alpha\sim5^\circ$.

\subsection{Comparison}

Finally, we compare the nanotube's voltage response for the two magnetic field orientations, $\mathbf{B} \parallel OX$ and $\mathbf{B} \perp OX$, see 
Fig.\,\ref{fig:1}(c), with those for the nanopetal and the flat membrane, both exposed to $\mathbf{B} \perp OX$ of the same magnitude, see Fig.\,\ref{fig:1}(a,b).

\begin{figure*}[t]
    \centering
    \includegraphics[width=\textwidth]{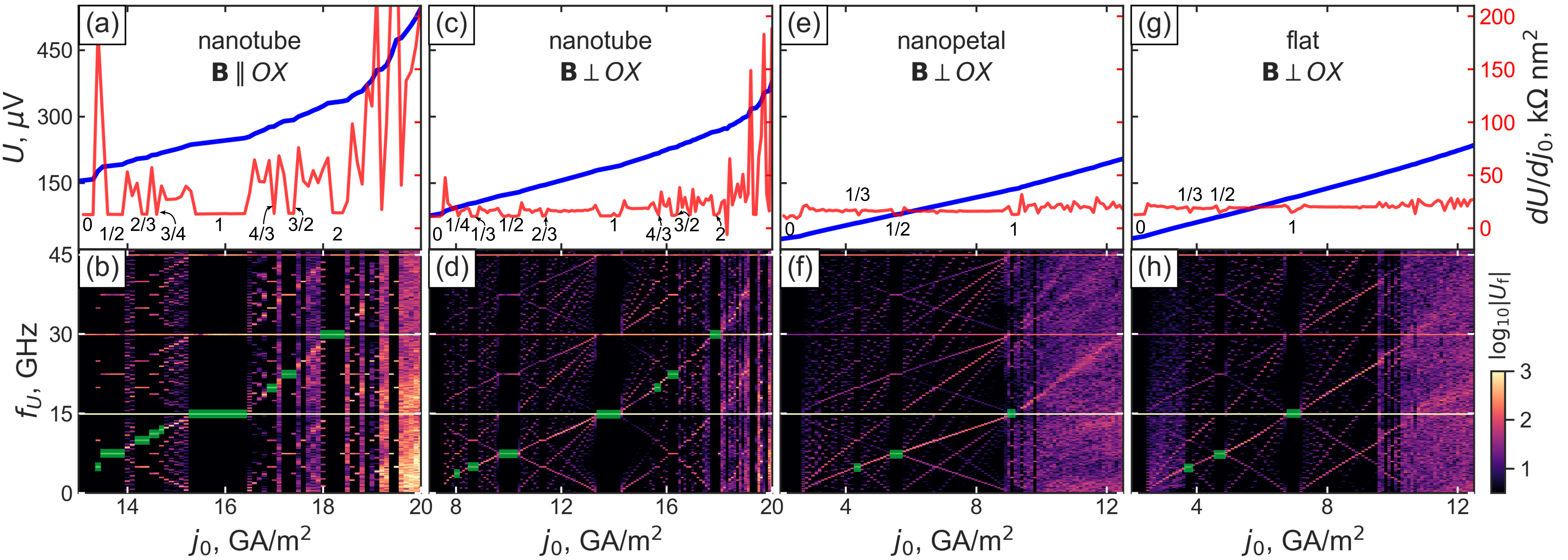}
    \caption{(a,c,e,g) The average voltage $U$ (blue) with its derivative $dU/dj_0$ (red) and (b,d,f,h) the voltage spectrum $f_\text{U}$ as a function of the dc transport current density $j_0$ for (a,b,c,d) the nanotube, (e,f) the nanopetal, and (g,h) the flat membrane. The magnetic field is oriented perpendicularly to the substrate except the cases (a,b) where it is parallel. The modulation frequency $f_\mathrm{j}=15$\,GHz, the modulation amplitude $j_1=3.0$ GA/m$^2$, and the magnetic field $B=5.2$ mT. The green lines highlight the nucleation frequency $f_1$ in the frequency-locking regime.}
    \label{fig:7}
\end{figure*}

In Fig.\,\ref{fig:7}, we compare the $I$-$V$ curves and the voltage spectra for the nanotube with different field orientations and with the nanopetal and the flat membrane. The width of the Shapiro steps and the height of the minima in the derivative $dU/dj_0$, Fig.\,\ref{fig:7}(a,c,e,g), increase in the following sequence: flat membrane, nanopetal, nanotube with $\mathbf{B}\perp OX$, and nanotube with $\mathbf{B}\parallel OX$. Fractional Shapiro steps are present even in the flat membrane. However, Shapiro steps in the flat membrane and the nanopetal are  very small and are thus very unlikely to be observed experimentally. In contrast, under otherwise the same conditions, the nanotube exhibits pronounced Shapiro steps, particularly for the $\mathbf{B} \parallel OX$ geometry. The absence of pronounced Shapiro steps in the nanopetal suggests that its radius is too large to confine vortices into a quasi-1D motion regime, which is in line with the spatial plot of the magnitude of the superconducting order parameter in Fig.\,\ref{fig:2}(b). We have also checked (not shown) that for a transition from the 2D regime to a quasi-1D vortex motion, the nanopetal radius should be decreased from $800$\,nm to about $400$\,nm.

\section{Discussion}
The major finding of the performed analysis is the revealed decisive impact of the change in dimensionality of the vortex motion on the appearance of the FLE. Without artificial pinning site arrays or notches facilitating the vortex penetration and motion along predefined paths, alone the strongly nonuniform normal component of the magnetic field allows for the realization of a quasi-1D vortex motion in 3D open nanotubes and suggests them as systems for the observation of the FLE. This qualitatively differs from planar systems where vortex dynamics occur in a 2D regime\,\cite{che14apl,wor12prb,ali20mic,dob15apl,wan24pra,pat21prb,jan06prb,lar17pra}. Moreover, a prediction is made regarding the FLE robustness upon deviation of the magnetic field direction by a few degrees away from the normal to the sample substrate.

We are now entering a discussion of the appearance of Shapiro steps in the studied structures and their features distinct from the previous works. In general, a fractional Shapiro step $n=p/q$ signalizes the presence of the FLE when $p$ vortices in each vortex chain nucleate every $q$ cycles of the stimulus modulation. Here,  $q$ is not divisible by $p$ and they are mutually prime. For example, the dynamics of vortices reflected in the fractional Shapiro step $n=2/3$  at the modulation frequency of $15$\,GHz can be described as three modulation cycles and two nucleation periods within $200$\,ps. Namely, the vortex nucleation occurs at the maximum of the ac transport current. Two vortices nucleate during the first two modulation cycles. However, during the third ac cycle, the nucleation of the third vortex is missing, as the membrane becomes overpopulated with the vortices near the edge where the nucleation should take place. In contrast, in the case of integer Shapiro steps, $n$ vortices nucleate within a single modulation cycle. As shown in Fig.\,\ref{fig:4}, fractional Shapiro steps are narrower than integer steps, indicating that fractional Shapiro steps require a more stable order in vortex chains than the integer steps.

The revealed integer and fractional Shapiro steps are expected to follow the same rules as Shapiro steps in other superconducting systems, since the magnetic flux of the vortex is quantized, sharing common properties across all superconducting systems. In general, the voltage corresponding to the $n$-th fractional or integer Shapiro step is given by the expression $U_0=n f_\mathrm{j} \Phi_0$, where ${\Phi_0 = h/(2e) \simeq 2.068\times10^{-15}\,\text{Wb}}$ is the magnetic flux quantum\,\cite{din08apl}. In the case of nanotubes, this expression should be multiplied by the number $n_\mathrm{c}=2$ of vortex chains moving in each half-tube, yielding
\begin{equation}
    \label{un}
    U_n = n_\mathrm{c} n f_\mathrm{j} \Phi_0.
\end{equation}
Equation \eqref{un} is consistent with another expression for the Shapiro step voltage $U_n = N k \left\langle v \right\rangle \Phi_0 / d$ in superconducting films with periodic antidot lattices\, \cite{van99prb}. Here, $N$ is the number of antidot rows, $k$ is the number of moving vortices in each lattice cell, $\left\langle v \right\rangle$ is the average velocity of vortices, and $d$ is the period of the antidot lattice. We consider that there is only one effective cell in each half-tube of the nanotube, since the tube edge controls the vortex nucleation in an open tube. Then, we obtain $d=L$ and $N=n_\mathrm{c}=2$ and the average velocity $\left\langle v \right\rangle$ and the number of vortices in each half-tube $k$ can be estimated as 
\begin{equation}
    \left\langle v \right\rangle = L/t_2,\qquad k = t_2 / t_1. 
\end{equation}
Here, $t_1$ is the period of vortex nucleation in the half-tube and $t_2$ is the time-of-flight for each vortex along the half-tube.
Then, the expression adapted from Ref.\,\cite{van99prb} to the nanotube coincides with Eq.\,\eqref{un} if we substitute the step order number $n$ by the ratio of the frequencies $n = f_1/f_\mathrm{j}$. Following Eq.\,\eqref{un}, the step voltage can be estimated as $U_0 = U_n / n \approx 62.1\,\mu\text{V}$, which is fairly close to the value  $U_0\approx 52.9\,\mu\text{V}$ obtained from Fig.\,\ref{fig:4}. The $\sim 15\%$ difference can be attributed to the fact that the theoretical estimation applies to flat membranes, whereas a corresponding expression for the nanotube is not available.

Finally, the experimental observability of integer and fractional Shapiro steps in open nanotubes depends on their stability under various perturbations that influence the vortex motion. Among these, lattice defects in nanotubes fabricated by particle beam-induced deposition or rolling-up technologies may affect the vortex motion, causing deviations from a straight line. However, since the vortex motion in nanotubes is quasi-1D, we expect Shapiro steps to demonstrate robustness against such defects, preserving the ordering of vortices even if their trajectories are distorted.

\section{Conclusions}
We have studied numerically the (dc+ac)-driven dynamics of vortices in superconductor open nanotubes in comparison with nanopetals and flat membranes. The frequency-locking effect and Shapiro steps have been predicted to be much pronounced for the open nanotubes and stable against deviations of the direction of the applied magnetic field from the normal to the substrate plane by a few degrees. The pronounced Shapiro steps are attributed to a reduction in the dimensionality of the moving vortex ensemble, evolving from the 2D vortex dynamics in planar membranes to two quasi-1D vortex chains moving in both half-cylinders of the open nanotube.  

The frequency-locking effect allows for tuning of the vortex nucleation frequency and average voltage via current modulation, and it can be used for voltage stabilization under perturbations that cause variations in the nucleation frequency. Therefore, integer and fractional Shapiro steps in the current-voltage curves of the open nanotubes provide a valuable tool for unveiling the vortex dynamics, which is challenging to visualize using local-probe techniques. Our predictions for the average voltage and its frequency spectrum can be examined experimentally, for example, on open Nb nanotubes fabricated by the self-rolling technology\,\cite{Loe19acs}. In all, our findings are relevant for superconducting devices, where vortex nucleation frequency and voltage stabilization by an external ac stimulus can enhance their operation.

\acknowledgments
The work of I.B. was funded by the Deutsche Forschungsgemeinschaft (DFG, German Research Foundation) under Germany’s Excellence Strategy -- EXC-2123 QuantumFrontiers -- 390837967. I.B. gratefully acknowledges the use of the CryoMind simulation workstation at CryoQuant/TU Braunschweig. V.M.F. gratefully acknowledges NHR Center NHR@TUD for the provided computing time. The authors are grateful for the support by the European Cooperation in Science and Technology (E-COST) Action CA21144. Further, O.D. acknowledges the E-COST for support via Grant No. E-COST-GRANT-CA21144-5883b676 and V.M.F. acknowledges the E-COST for support via Grant No. E-COST-GRANT-CA21144-4144281a.

\appendix
\counterwithin{equation}{section}
\section{TDGL modeling}
\label{app:model}

The numerical simulations were done relying upon the TDGL equation. In the dimensionless form, the TDGL reads
\begin{equation}
\label{eGL}
    (\partial_t + i \varphi)\psi=\left(\nabla
    -i\textbf{A}\right)^{2}\psi
    +\left(1-|\psi|^{2}\right)\psi,
\end{equation}
where $\psi$ is the order parameter, $\varphi$ is the scalar potential associated with the electric field $E=-\nabla\varphi$, and $\mathbf{A}$ is the vector potential tangent to the surface, which describes the external normal-to-surface magnetic field $\mathbf{B}_n = [\nabla \times \mathbf{A}]$. The dimensional units are reported in Table\,\ref{tab:units}. The thickness of the membrane is considered to be small enough ($50$\,nm) to neglect the magnetic field induced by the superconducting and normal currents. The scalar potential obeys the Poisson equation
\begin{equation}\label{poisson}
    \Delta \varphi = \sigma_\text{n}^{-1} \nabla \cdot \mathbf{j}_\text{sc},\qquad
    \mathbf{j}_\text{sc} = \Im \left(
        \psi^* (\nabla - i \mathbf{A}) \psi
    \right),
\end{equation}
where $\sigma_\text{n}$ is the dimensionless normal conductivity. 

The boundary conditions for the order parameter and scalar potential read
\begin{subequations}
\label{eq:boundaries}
\begin{align}
    \psi=0,\quad
    \partial_x \varphi = - j_\mathrm{tr} / \sigma_\mathrm{n},
    \qquad
    \text{at contacts } \partial D_x,
    \\
    (\partial_y - i A_y) \psi = 0,\quad
    \partial_y \varphi = 0\quad
    \text{at free edges } \partial D_y.
\end{align}
\end{subequations}
Link variables are applied to reduce the error associated with the gauge invariance of the system. The parameters used for the simulations are listed in Table\,\ref{table:parameters}. The voltage is calculated as the difference of the scalar potential averaged over the arc-length $L$. The grid resolution is set to $192 \times 384$\,points, with a minimum of $600 000$ time steps, each spanning $25$\,fs.
\begin{table}[ht]
\centering
\caption{Dimensional units for the quantities in Eqs.\,(\ref{eGL})--(\ref{poisson}) for Nb at $T/T_c = 0.952$. 
\label{tab:units}}
\begin{tabular}{P{3.5cm} P{2cm} P{2.5cm} }
 \hline
 \textbf{Parameter} & \textbf{Unit} & \textbf{Value} \\
 \midrule
 Time & $\xi^2/D$ & 2.8 ps
 \\
 Length & $\xi$ & 60 nm
 \\
 Magnetic field & $\Phi_0 / 2\pi\xi^2$ & 92 mT
 \\
 Current density & $\hslash c^2 / 8\pi\lambda^2 \xi e$ & 60 GA $\text{m}^{-2}$
 \\
 Electric potential & $\sqrt{2}H_c \xi \lambda / c \tau $ & 111 $\mu$V
 \\
 Conductivity & $c^2 / 4\pi\kappa^2D$ & 31 $(\mu\Omega \text{ m})^{-1}$
 \\
 \midrule
 \bottomrule
\end{tabular}
\end{table}

\begin{table}[ht]
    \centering
    \caption{Material and geometric parameters of the Nb nanoarchitectures used in the TDGL simulations.}
    \begin{tabular}{P{3.5cm} P{2cm} P{2.5cm}}
    \hline
    \textbf{Parameters} & \textbf{Denotation} & \textbf{Value for Nb}\\
    \midrule
    Electron mean free path & $l$ & 6 nm \\
    Fermi velocity&$v_F$ &$6\times10^{-5}$\,m/s \\
    Diffusion coefficient &$D$ &$1.2\times10^{-3}$\,m$^2$/s \\
    Normal conductivity &$\sigma_\mathrm{n}$ &$16\,(\mu\Omega\mathrm{ m})^{-1}$ \\
    Relative Temperature & $T/T_c$ & 0.952\\
    Penetration depth&$\lambda$ &278 nm \\
    Coherence length&$\xi$ &60 nm \\
    GL parameter&$\kappa$ &4.7 \\
    \midrule
    \bottomrule
    \end{tabular}
    \label{table:parameters}
 \end{table}

\FloatBarrier

\bibliography{main}

\end{document}